\documentclass[12pt,oneside]{article}
\usepackage{amsfonts,amssymb,graphicx}
\def\be{\begin{equation}}
\def\ee{\end{equation}}
\def\bea{\begin{eqnarray}}
\def\eea{\end{eqnarray}}
\def\<{\langle}
\def\>{\rangle}
\def\~{\tilde}
\def\s{\sigma}
\def\l{\lambda}

\def\b{\beta}

\def\t{\tau}

\def\ud{\mathrel{\mathop{=}\limits^{\cal D}}}

\newcommand{\B}{\Bbb B}

\newcommand{\Z}{\Bbb Z}

\newcommand{\av}[1]{\mbox{{\rm Av}}\left(#1\right)}

\newtheorem{theorem}{Theorem}

\begin{document}
\begin{center}
{\bf\sc\Large spin-glass stochastic stability:\\
a rigorous proof.}\\
\vspace{1cm}
{Pierluigi Contucci, Cristian Giardin\`a}\\
\vspace{.5cm}
{\small Dipartimento di Matematica} \\
    {\small Universit\`a di Bologna,
    40127 Bologna, Italy}\\
    {\small {e-mail: {\em contucci@dm.unibo.it, giardina@dm.unibo.it}}}\\
\vskip 1truecm
%{\small DATE}
\end{center}
\vskip 1truecm
\begin{abstract}\noindent
We prove the property of stochastic stability previously introduced
as a consequence of the (unproved) continuity hypothesis in the temperature
of the spin-glass 
quenched state. We show that stochastic stability  holds in $\beta$-average
for both the
Sherrington-Kirkpatrick model in terms of the square of the overlap function
and for the Edwards-Anderson model in terms of the bond overlap. We show
that 
the volume rate at which the property is reached in the thermodynamic limit
is $V^{-1}$. As a byproduct we show that the stochastic stability identities
coincide with those obtained with a different method by Ghirlanda and Guerra
when applyed to the thermal fluctuations only.

\end{abstract}
\newpage\noindent
\section{Introduction}
In a previous paper by Aizenman and Contucci \cite{AC} the property of
stochastic stability
was introduced as the consequence of a continuity (in term of the inverse
temperature $\beta$)
hypothesis of the
quenched state for the Sherrington-Kirkpatrick \cite{SK} model.
Stochastic stability says that a suitable class of perturbations of the spin
glass Hamiltonian
produces very small changes in the quenched equilibrium state and that such
a change 
vanishes in the thermodynamic limit. This property has interesting
consequences for the spin glass models: in terms of the overlap distribution
it implies that  
the quenched measure is {\it replica-equivalent} \cite{MPV,P} a property
originally introduced
within the replica symmetry breaking Parisi ansatz. The same property is
also used in 
\cite{FMPP1,FMPP2} to build a bridge between  equilibrium and
off-equilibrium properties
in a spin-glass model being these last the only ones physically accessible
to experimental 
investigation. More recently all and only the constraints that stochastic
stability implies 
for the overlap moments have been completely classified \cite{C,BCK}.

In this paper we give
a rigorous proof of stochastic stability property in {\it $\beta$-average}.
This result is achieved in an elementary way by use of the sum law
for independent Gaussian variables and works in full generality 
for both mean-field and finite dimensional spin glass models. 
We also derive the explicit form of the stochastic stability identities
which first appeared in \cite{AC}
and we prove, using integration by parts in the spirit of
\cite{CDGG}, that they coincide
with a subset of the Ghirlanda-Guerra identities \cite{G,GG}, namely 
the part related to the thermal fluctuation bound (see also \cite{T}
for a nice set of rigorous results derived from those identities). 

The  proof also provides the rate at which
stochastic stability in $\beta$-average is reached
with the thermodynamic limit  which turns out to be $V^{-1}$. 
The paper is
organized with Sec. 2 containing  a list of the definitions and the
statement of the two main
theorems. Their proof is built in Sec. 3 while Sec. 4 shows how
to apply the results to
both the mean field models, which we illustrate for the
Sherrington-Kirkpatrick model \cite{SK}, and for the
finite dimensional cases with the  Edwards-Anderson model \cite{EA}. Sec. 5
collects some comments.

\section{Definitions and Results}

We consider a disordered model of Ising configurations
$\s_n=\pm 1$, $n\in \Lambda\subset \Z^d$ for some $d$-parallelepiped
$\Lambda$ of volume $|\Lambda|$. We denote
$\Sigma_\Lambda$ the set of all $\s=\{\s_n\}_{n\in \Lambda}$, and
$|\Sigma_\Lambda|=2^{|\Lambda|}$. In the sequel the
following definitions will be used.

\begin{enumerate}

\item Hamiltonian. For every $\Lambda\subset \Z^d$ let
$\{H_\Lambda(\sigma)\}_{\s\in\Sigma_N}$
be a family of
$2^{|\Lambda|}$ {\em translation invariant (in distribution) centered
Gaussian} random variables
of {\em volume-size} covariance matrix
\be\label{cc}
\av{H_\Lambda(\s) H_\Lambda (\tau)} \; = \; \,|\Lambda| \,{\cal Q}_\Lambda
(\s,\tau) \, ,
\ee
and
\be
{\cal Q}_\Lambda (\s,\s) \; = \;  1 \, .
\ee
By the Schwarz inequality $|{\cal Q}_\Lambda (\s,\t)|\le 1$ for all $\s$ and
$\t$. 
\item Random partition function
\be
{\cal Z}(\beta) \; := \; \sum_{\s  \in \,\Sigma_\Lambda}
e^{-\beta{H}_\Lambda(\s)}
\; .
\ee
\item Random free energy ${\cal F}(\beta)$
\be
-\beta {\cal F}(\beta) \; := \; {\cal A}(\beta) \; := \; \ln {\cal Z}(\beta)
\; .
\ee
\item Quenched free energy $F(\beta)$
\be
-\beta F(\beta) \; := \; A(\beta) \; := \; \av{ {\cal A}(\beta) }\; .
\ee
%\be 
%F_\Lambda(\beta) \; := \; \av{{\cal F}_\Lambda(\beta)} \; .
%\label{fe}
%\ee
\item
$R$-product random Gibbs-Boltzmann state
\be
\Omega (-) \; := \;
\sum_{\sigma^{(1)},...,\sigma^{(R)}}(-)\,
\frac{
e^{-\beta[H_\Lambda(\s^{(1)})+\cdots
+H_\Lambda(\sigma^{(R)})]}}{[{\cal Z}(\beta)]^R}
\; .
\label{omega}
\ee
\item Quenched equilibrium state
\be
<-> \, := \av{\Omega (-)} \; .
\ee
\item\label{obs} Observables. For any smooth bounded function $G({\cal
Q}_{\Lambda})$ 
(without loss of generality we consider $|G|\le 1$)
of the covariance matrix entries we introduce the random (with respect to
$<->$) $R\times R$ matrix $Q=\{q_{k,l}\}$
by the formula
\be
<G(Q)> \; := \; \av{\Omega (G({\cal Q}_{\Lambda}))} \; .
\ee
E.g.: $G({\cal Q}_\Lambda)={\cal
Q}_{\Lambda}(\sigma^{(1)},\sigma^{(2)}){\cal
Q}_{\Lambda}(\sigma^{(2)},\sigma^{(3)})$
\be
<q_{1,2}q_{2,3}> \; = \;
\av{\frac{\sum_{\sigma^{(1)},\sigma^{(2)},\sigma^{(3)}}{\cal
Q}_{\Lambda}(\sigma^{(1)},\sigma^{(2)}){\cal
Q}_{\Lambda}(\sigma^{(2)},\sigma^{(3)})\,e^{-\beta[\sum_{i=1}^{3}H_\Lambda(\
s^{(i)})]}}{[{\cal Z}(\beta)]^3}}
\ee
\item \label{deform}
Deformed quenched state. For every $\Lambda\subset \Z^d$ let the
$\{K_\Lambda(\sigma)\}_{\s\in\Sigma_N}$ be a {\em translation invariant
centered Gaussian} random family of {\em size one} covariance matrix
\be
\av{K_\Lambda(\s) K_\Lambda (\tau)} \; = \;  {\cal Q}_\Lambda (\s,\tau) \, ,
\ee
where the families $H$ and $K$ are mutually independent with respect to the
joint Gaussian distribution, i.e.
\be
\av{H_{\Lambda}(\s)K_\Lambda(\t)}\; = \; 0 \; .
\ee
We consider
\be
{\cal Z_\lambda(\beta)} \; := \; \sum_{\s\in\,\Sigma_{\Lambda}}e^{- \beta
H_\Lambda(\s)+\sqrt{\l}K_\Lambda(\s)} \; ,
\ee

\be
A_\lambda(\beta) \; := \; \av{\ln {\cal Z_\lambda(\beta)}} \; ,
\ee

\be
\Omega_{\lambda}(-) \; := \; \frac{\Omega ( (-)\,
e^{\sqrt{\lambda} \,[K_\Lambda(\sigma^{(1)}) + \cdots +
K_\Lambda(\sigma^{(R)})]})}
{\Omega (e^{\sqrt{\lambda} \,[ K_\Lambda(\sigma^{(1)})
+ \cdots + K_\Lambda(\sigma^{(R)}) ]})} \; ,
\ee

and the deformed quenched state
\be
<->_{\lambda} \, := \, \av{\Omega_{\lambda}(-)} \; .
\ee
\item Stochastic Stability.
The quenched measure is said to be {\it stochastically stable} if for
every observable $G$ (see def.
\ref{obs}) the deformed state is stationary in the thermodynamic limit:
\be
\lim_{\Lambda\,\nearrow\,\Z^d} \frac{d}{d\lambda}<G>_\l \; = \; 0
\ee
It is possible to see (within Theorem \ref{expli}) that there is a
function of the overlap matrix
elements: $\Delta G$ s.t.
\be
\label{dg}
<\Delta G>_\lambda  \, := \, \frac{d}{d\lambda}<G>_{\lambda} \; .
\ee
A stochastically stable measure fulfills then the property
\be
\lim_{\Lambda\,\nearrow\,\Z^d} <\Delta G>_{\lambda} \; = \; 0
\ee
for all the observables $G$.
\end{enumerate}
Our main result state that a spin glass model is stochastically stable
$\b$-{\it almost everywhere}
(Theorem 1), characterizes the functions $\Delta G$ (Theorem 2) and
establish their coincidence with
the quantities obtained with the Ghirlanda-Guerra method when applied only
to the thermal fluctuations.

\begin{theorem}[Stochastic Stability]
\label{main}
The spin-glass quenched state is stochastically stable {\rm in
$\beta$-average}, i.e.
for each interval $[\beta_1,\beta_2]$  and each observable
$G$ (as in def. \ref{obs}):
\be
\left|\int_{\beta^2_1}^{\beta^2_2} <\Delta G>_{\lambda} \; d\beta^2 \right|
\; \le \; 
\frac{2}{|\Lambda|} \; .
\ee
\end{theorem}
\begin{theorem}[Zero average Observables]
\label{expli} 
The explicit form of the zero average quantities is
\be
2 \Delta G \; =\; 
\mathop{\sum_{k,l=1}^{R}}_{k\ne l} G \,q_{\,l,\,k}
- 2R G \, \sum_{l=1}^{R} q_{\,l,\,R+1}
+R(R+1) G \, q_{\,R+1,\,R+2} \; ,
\ee
which coincide with thermal part of the Ghirlanda-Guerra identities.
\end{theorem}

\section{Proof of the results}

\noindent

\vspace{0.3cm}
\noindent
{\bf Proof of Theorem \ref{main}:}
Since for $\widetilde H$ independent from $H$ and $K$ and distributed like
$H$ we have,
in distribution, that
\be
-\beta H_{\Lambda}  +
\sqrt{\lambda} K_{\Lambda}  \ud
-\sqrt{\beta^2 + \frac{\lambda}{|\Lambda|}} \,{\widetilde H}_{\Lambda}
\ee
from the def. (\ref{deform}) of the deformed quenched state
of the function $G$, all the expectations $<G>_\lambda$ turn out to be
functions of
$\beta^2 + \frac{\lambda}{|\Lambda|}$: $<G>_\lambda=g\left(\beta^2 +
\frac{\lambda}{|\Lambda|}\right)$. From the composite function derivation
rule we deduce (the prime denotes
derivative w.r.t. the argument):
\be
\frac{d}{d\lambda}<G>_\lambda \; = \; g' \left(\beta^2 +
\frac{\lambda}{|\Lambda|}\right)\cdot
\frac{1}{|\Lambda|}
\ee
and
\be
\frac{d}{d\beta}<G>_\lambda \; = \; g' \left(\beta^2 +
\frac{\lambda}{|\Lambda|}\right)\cdot 2\beta \; ,
\ee
from which we have
\be
2\b \frac{d}{d\lambda}<G>_\lambda  \; = \;
\frac{1}{|\Lambda|}\frac{d}{d\beta}<G>_\lambda \; .
\label{eqq}
\ee
Integrating in $d\b$ and using the foundamental theorem
of calculus we obtain
\be
\int_{\beta^2_1}^{\beta^2_2} <\Delta G>_{\lambda} \; d\beta^2 =
\frac{<G>_{\lambda}(\beta_2) - <G>_{\lambda}(\beta_1)}{|\Lambda|}
\ee
Remembering the assumption on boundedness of function $G$
(def. \ref{obs}) this complete the proof.

\vspace{0.3cm}
\noindent
{\bf Proof of Theorem \ref{expli}:} let $h(\s)=|\Lambda|^{-1}H_\Lambda(\s)$ be the
Hamiltonian per particle. From formula (\ref{eqq}) and a direct computation of the derivative
of $<G>_{\lambda}$ with respect to the inverse temperature we have 
\be
\label{iden}
- 2 \beta <\Delta G>_{\lambda}\; = \;
\sum_{l=1}^R
\av{\Omega_\lambda(h(\s^{(l)})\,G)-\Omega_\lambda(h(\s^{(l)}))
\Omega_\lambda(G)} \;.
\ee
For each replica $l$ $(1\le l\le R)$, we evaluate separatly
the two terms in the right side of Eq. (\ref{iden}) by using
the integration by parts (generalized Wick formula) for
correlated Gaussian random variables, $x_1,x_2,\ldots,x_n$
\be
\av { x_i\, \psi(x_1,...,x_n) } = 
\sum_{j=1}^n \av { x_i x_j  } \,  
\av {\frac{\partial \psi(x_1,...,x_n)}{\partial x_j} } \;.
\ee
It is convenient to denote by $p_{\lambda}\,(R)$ the Gibbs-Boltzmann weight
of R copies of the deformed system
\be
p_{\lambda}\,(R) \,= \,
\frac{
e^{-\beta\,[\,\sum_{k=1}^R H_\Lambda(\s^{(k)})\,]\, +
\;\sqrt{\lambda}\,[\,\sum_{k=1}^R K_\Lambda(\s^{(k)})\,]\,}}
{[{\cal Z}_\lambda(\beta)]^R} \;,
\ee
so that we have
\be
\label{derivBolt}
- \frac{1}{\beta}\frac{dp_{\lambda}\,(R)}{dH_{\Lambda}(\tau)} \;=\;
p_{\lambda}\,(R) 
\left(\sum_{k=1}^R \delta_{\s^{(k)},\,\tau}\right)
- R \;p_{\lambda}\,(R)\;
\frac{
e^{-\beta[H_\Lambda(\tau)]}}
{[{\cal Z}_\lambda(\beta)]} \;.
\ee
We obtain
\bea
\av{\Omega_\lambda(h(\s^{(l)})\,G)} & = &
\frac{1}{|\Lambda|}\,
\av{\;
\sum_{\sigma^{(1)},...,\sigma^{(r)}}\;
G\;H_{\Lambda}(\s^{(l)}) \;
p_{\lambda}\,(R)} \\
& = & 
\label{part1}
\av{ \;
\sum_{\sigma^{(1)},...,\sigma^{(r)}}\;\sum_{\tau}\;G\;
{\cal Q}_{\Lambda}(\s^{(l)},\tau)\;
\frac{dp_{\lambda}\,(R)}{dH_{\Lambda}(\tau)}}
\qquad\qquad
\\
& = & 
\label{temp1}
- \beta\,\left [
 <G>_\lambda + \mathop{\sum_{k=1}^{R}}_{k\ne l} <G \,q_{\,l,\,k}>_\lambda -
R  <G \, q_{\,l,\,R+1}>_\lambda
\right]
\eea
where in (\ref{part1}) we made use of the integration
by parts formula and (\ref{temp1}) is obtained
by (\ref{derivBolt}).
Analogously, the other term reads
\bea
\;
\av{\Omega_{\lambda}(h(\s^{(l)}))\,\Omega_{\lambda} (G)} & = &
\frac{1}{|\Lambda|}\,
\av{\;
\sum_{\sigma^{(l)}}\sum_{\tau^{(1)},...,\tau^{(R)}}\;
G\;H_{\Lambda}(\s^{(l)}) \;
p_{\lambda}\,(R+1)} \\
& = & 
\label{part2}
\av{\;
\sum_{\sigma^{(l)}}\sum_{\tau^{(1)},...,\tau^{(R)}}\;\sum_{\gamma}\;G\;
{\cal Q}_{\Lambda}(\s^{(l)},\gamma)\;
\frac{dp_{\lambda}\,(R+1)}{dH_{\Lambda}(\gamma)}}\quad
\qquad
\\
& = & 
\label{temp2}
- \beta\,\left [
 <G>_\lambda + R <G \, q_{\,l\,R+1}>_\lambda
- (R+1) <G \, q_{\,R+1,\,R+2}>_\lambda
\right]\nonumber\\
\eea
Inserting the (\ref{temp1}) and (\ref{temp2}) in Eq. (\ref{iden})
we finally obtain Theorem \ref{expli}.

\noindent
{\bf Remark:} The proof of the Theorems shows that the identities
which follow from the stochastic stability property are  
included in the Ghirlanda-Guerra identities \cite{GG}.
Indeed the family of GG identities are obtained  
from the self-averaging of the internal energy per particle with
respect to the full equilibrium quenched measure.
This implies, by the use of the Cauchy-Schwartz inequality, the
vanishing of the truncated correlation between
internal energy per particle and a generic observable $G$
in the thermodynamic limit:
\be
<h G> - <h><G> \rightarrow 0 \quad\quad as \quad
|\Lambda| \rightarrow \infty \; .
\ee
But clearly the previous fluctuation can be decomposed 
as a sum of the thermal fluctuation (averaged over the Gaussian
disorder) and the fluctuation with respect to the disorder itself, 
{\it i.e.}
\bea
<h \; G> - <h><G>  
& = & 
\av{\Omega[h\,G]} - \av{\Omega[h]} \av{\Omega[G]}  \nonumber \\
& = &
\label{pippo}
\av{\Omega[h\,G] - \Omega[h]\Omega[G])}  \,+ \\
&   & 
\av{\Omega[h]\Omega[G]} - \av{\Omega[h]} \av{\Omega[G]}
\eea
By formula (\ref{iden}) we see that the thermal 
fluctuations (Eq.(\ref{pippo})) are those controlled by the stochastic stability.

\section{Models}
The results proved in the previous sections hold true in complete generality
because they are based on the general property of Gaussian variables.
Stochastic stability  in particular is fulfilled by both mean
field
models (like the Sherrington-Kirkpatrick, its $p$-spin generalization, the
REM and
GREM models etc.) and by the finite dimensional models (like the
Edwards-Anderson and
Random Field models in general dimension $d$). The main point to be observed
and well stressed is that
each one of these models has his own set of observables which describe the
quenched
equilibrium state, namely the Gaussian covariance matrix of their own
Hamiltonians, see 
Eq. (\ref{cc}). To be more specific let illustrate the two main cases of the
covariance matrix 
for the Sherrington-Kirkpatrick model and for the Edwards-Anderson.
The SK model of Hamiltonian
\be
H_N(\s,J) \; = \; - \, \frac{1}{\sqrt{N}} \sum_{i,j=1}^N J_{i,j}\s_i\s_j
\ee
with $\{J_{ij}\}$  identical independent normal Gaussian variables
has a covariance matrix given by the standard overlap function between two
configurations:
\be\label{csk}
{\cal Q}^{(SK)}_{\Lambda}(\s,\tau) \; = \;
\left[\frac{1}{N}\sum_{i=1}^{N}\s_i\t_i\right]^2
\ee
The Edwards-Anderson Hamiltonian is
\be
H_{\Lambda}(J,\s) \, = \, - \sum_{(n,n')\in B(\Lambda)}J_{n,n'}\s_n\s_{n'}
\; ,
\label{eag}
\ee
where the $J_{n,n'}$ are again independent normal Gaussian variables and the
sum
runs over all  pairs of nearest neighbors sites $n,n' \in \Lambda \subset
\Z^d$ 
with $|n-n'|=1$. Using
the standard identification of
the space of nearest neighbors with the $d$-dimensional {\it bond}-lattice
$b\in \B^d$ with $b=(n,n')$
and denoting $B(\Lambda)$ the $d$-bond-parallelepiped associated to
$\Lambda$ 
$(|B| = d|V|)$ we
introduce, for two spin configurations   $\s$ and $\t$, the notation
$\s_b=\s_n\s_{n'}$ and $\t_b=\t_n\t_{n'}$.
The covariance matrix turns out to be
\be
{\cal Q}^{(EA)}_{\Lambda}(\s,\tau) := \frac{1}{|B|}\sum_{b\in B}
Q_b(\s,\t)\; .
\label{bov}
\ee
where the local bond-overlap $Q_b(\s,\t)$ between $\s$ and $\t$  is
\be
Q_b(\s,\t):=\s_b\t_b \; .
\ee
The property of stochastic stability for the Edwards-Anderson model in terms
of its
link-overlap has been originally considered in \cite{C2}. The theorem proved
here provides
the generalization to the generic observable $G$.

\section{Comments}
In this paper we have proved that every Gaussian spin glass model
is stochastically stable with respect to a suitable class of perturbations.
The consequences of such a stability can be expressed as zero average
observables in terms of the proper overlap that each model carries: the
covariance of its own Hamiltonian. 
It is finally worth to
mention that
the identities that we proved for the Edwards-Anderson model are compatible
with both the pictures of triviality and those of non-triviality for the
overlap 
distribution at low temperature; for a discussion the reader may see the
replica symmetry 
Breaking theory in \cite{MPV}, the Droplet theory in \cite{FH,BM}, the
chaotic theory in \cite{NS} and the
trivial-non-trivial in \cite{PY,KM}. Nevertheless the stochastic stability
identities could suggest a test of
triviality for the suitable overlap distribution in the same spirit of
\cite{MPRRZ}. We plan to return on these
questions in a future work.
\vskip .5truecm
\noindent
{\bf Acknowledgments}.
We thank F. Guerra for many interesting discussions and in particular
for an observation which led to a substantial improvement of this work.
We also thanks A. Bovier, A. van Enter, S. Graffi, M. Talagrand
and
F.L. Toninelli.

\end{document}